\documentclass[twocolumn,aps,showpacs,prl]{revtex4}
\usepackage{graphicx}
\usepackage{epstopdf}

\begin{document}

\title{Antiferromagnetic superexchange interactions in LaOFeAs}

\author{Fengjie Ma$^{1}$}
\author{Zhong-Yi Lu$^{2}$}\email{zlu@ruc.edu.cn}
\author{Tao Xiang$^{3,1}$}\email{txiang@aphy.iphy.ac.cn}

\date{\today}

\affiliation{$^{1}$Institute of Theoretical Physics, Chinese Academy
of Sciences, Beijing 100190, China }

\affiliation{$^{2}$Department of Physics, Renmin University of
China, Beijing 100872, China}

\affiliation{$^{3}$Institute of Physics, Chinese Academy of
Sciences, Beijing 100190, China }

\begin{abstract}

From first-principles calculations, we have studied the electronic
and magnetic structures of the ground state of LaOFeAs. The Fe spins
are found to be collinear antiferromagnetic ordered, resulting from
the interplay between the strong nearest and next-nearest neighbor
superexchange antiferromagnetic interactions. The structure
transition observed by neutron scattering is shown to be
magnetically driven. Our study suggests that the antiferromagnetic
fluctuation plays an important role in the Fe-based superconductors.
This sheds light on the understanding of the pairing mechanism in
these materials.

\end{abstract}


\maketitle


Recently an iron-based material LaOFeAs was reported to show
superconductivity with a transition temperature $T_c\sim 26K$ by
partial substitution of O with F atoms\cite{kamihara}. Soon after,
other families of Fe-As oxyarsenides with La replaced by
Sm\cite{xhchen}, Ce\cite{chen3}, Pr\cite{ren2} and other rare earth
elements were found superconducting with $T_c$ more than 50K. Like
cuprates, these iron arsenides have a layered structure. The
superconducting pairing is believed to happen in the iron-based FeAs
layers. The high transition temperature and the preliminary band
structure calculation suggests that the superconductivity in these
Fe-arsenide superconductors is not mediated by electron-phonon
interaction. It is commonly believed that the understanding of
electronic structures of the parent compound LaOFeAs is the key to
determine the underlying mechanism to make it superconducting upon
doping.

The early band structure calculations suggested that the pure
LaOFeAs compound is a nonmagnetic metal but with strong
ferromagnetic or antiferromangtic (AFM) instability
\cite{singh,xu,kotliar}. Later, it was found that the
antiferromagnetically ordered state\cite{ma,ccao} has a lower energy
than the nonmagnetic one, probably due to the Fermi surface
nesting\cite{ma}. Dong et al.\cite{dong} predicted that the AFM
state should form a collinear-striped structure by breaking the
rotational symmetry. This collinear ordered AFM state has indeed
been observed by the neutron scattering experiment
\cite{cruz,mcguire}. Furthermore, the neutron scattering measurement
found that there is a structure transition with a monoclinic lattice
distortion at $\sim 150$K and the collinear order is formed about
15$\sim$20K below this transition temperature. Without this
distortion, the square AFM order induced purely by the Fermi surface
nesting is expected to be more stable since there are two orthogonal
but equivalent nesting directions ($\pi,~\pi$) and ($\pi,~-\pi$),
which can lower the energy of the ground state by keeping its
rotational symmetry\cite{ma}.

In this paper, we report the theoretical result on the electronic
and magnetic structures of the ground state of LaOFeAs obtained from
first-principles band structure calculations. We find that there are
strong nearest and next-nearest neighbor superexchange interactions
in this material (similar conclusion was obtained by
Yildirim\cite{yild}). The nearest and next nearest neighbor
superexchange interactions have almost the same amplitude within
error of calculation. Their competition affects strongly the
electronic structure of the ground state. This drives a small
monoclinic lattice distortion and a collinear ordering of Fe spins,
as observed by neutron scattering.

In our calculations the plane wave basis method was
used\cite{pwscf}. We adopted the local (spin) density approximation
and the generalized gradient approximation of
Perdew-Burke-Ernzerhof\cite{pbe} for the exchange-correlation
potentials. The ultrasoft pseudopotentials \cite{vanderbilt} were
used to model the electron-ion interactions. After the full
convergence test, the kinetic energy cut-off and the charge density
cut-off of the plane wave basis were chosen to be 600eV and 4800eV,
respectively. The Gaussian broadening technique was used and a mesh
of $16\times 16\times 8$ k-points were sampled for the
Brillouin-zone integration. LaOFeAs has a tetragonal layered
structure with $P4/nmm$ symmetry. A crystal unit cell consists of
eight atoms with alternating FeAs and LaO layers along the c axis.
In the calculation, the internal atomic coordinates within the cell
were determined by the energy minimization.

\begin{figure}
\includegraphics[width=6cm]{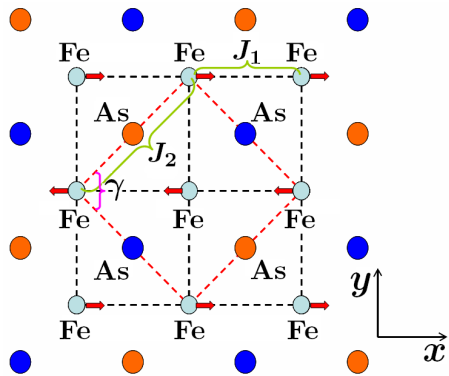}
\caption{(Color online) Schematic top view of the FeAs layer in
LaOFeAs. The small dashed square is an $a\times a$ unit cell while
the large dashed square is a $\sqrt{2}a\times \sqrt{2} a$ unit cell.
The collinear ordered Fe spins in the ground state are shown by red
arrows.} \label{figa}
\end{figure}

The previous band structure calculations
\cite{singh,xu,kotliar,ccao,ma} used an $a\times a\times c$ crystal
unit cell as the working cell, in which two Fe, As, La, and O atoms
were included. To explore the magnetic structure, in particular the
collinear AFM state of LaOFeAs, here we use a
$\sqrt{2}a\times\sqrt{2}a\times c$ unit cell (Fig. 1). In order to
determine the values of the nearest and next-nearest neighbor
coupling constants of spin-spin interaction, $J_1$ and $J_2$ (see
Fig. 1), we have evaluated the minimal energies of four different
magnetic states of Fe ions with constraints imposed if not stable.
These four states have nonmagnetic, ferromagnetic, square AFM, and
collinear AFM orders, respectively. If the energy of the nonmagnetic
state of LaOFeAs is set to zero, we find that the energies of the
ferromagnetic, square AFM, and colliear AFM states are (0.0905,
-0.010875, -0.21475) eV/Fe, respectively. Thus the ground state is a
collinear-ordered AFM state, in agreement with the experimental
observation\cite{cruz,mcguire}.

The magnetic moment around each Fe atom is found to be about
$2.2\sim 2.6~\mu_B$, varying weakly in the above three magnetically
ordered states. This suggests that the spin of Fe ions is between 1
and 3/2. The magnetic moment obtained from the neutron scattering
model is about 6 times smaller than this result. This is probably
because the correlated effect, especially the strong competition
between different AFM states, has not been fully included in the
density functional theory calculation.

To quantify the AFM interactions in this material, we assume that
these energy differences are predominantly contributed from the
interactions between the Fe spins which can be modeled by the
following frustrated Heisenberg model with the nearest and
next-nearest neighbor couplings $J_1$ and $J_2$
\begin{equation}\label{eq:Heisenberg}
H=J_1\sum_{\langle ij \rangle}\vec{S}_i\cdot\vec{S}_j +J_2\sum_{ \ll
ij \gg}\vec{S}_i\cdot\vec{S}_j,
\end{equation}
whereas $\langle ij \rangle$ and $\ll ij \gg$ denote the summation
over the nearest and next-nearest neighbors, respectively. From the
calculated energy data, we find that $J_1 \sim 0.0498eV/S^2$ and
$J_2 \sim 0.0510 eV/S^2$ (the detail how these parameters are
determined is given in the Appendix). If the spin of each Fe ion $S
= 1$, then $J_1 \sim 0.0498eV$ and $J_2 \sim 0.0510 eV$.

\begin{figure}
\includegraphics[width=8.5cm]{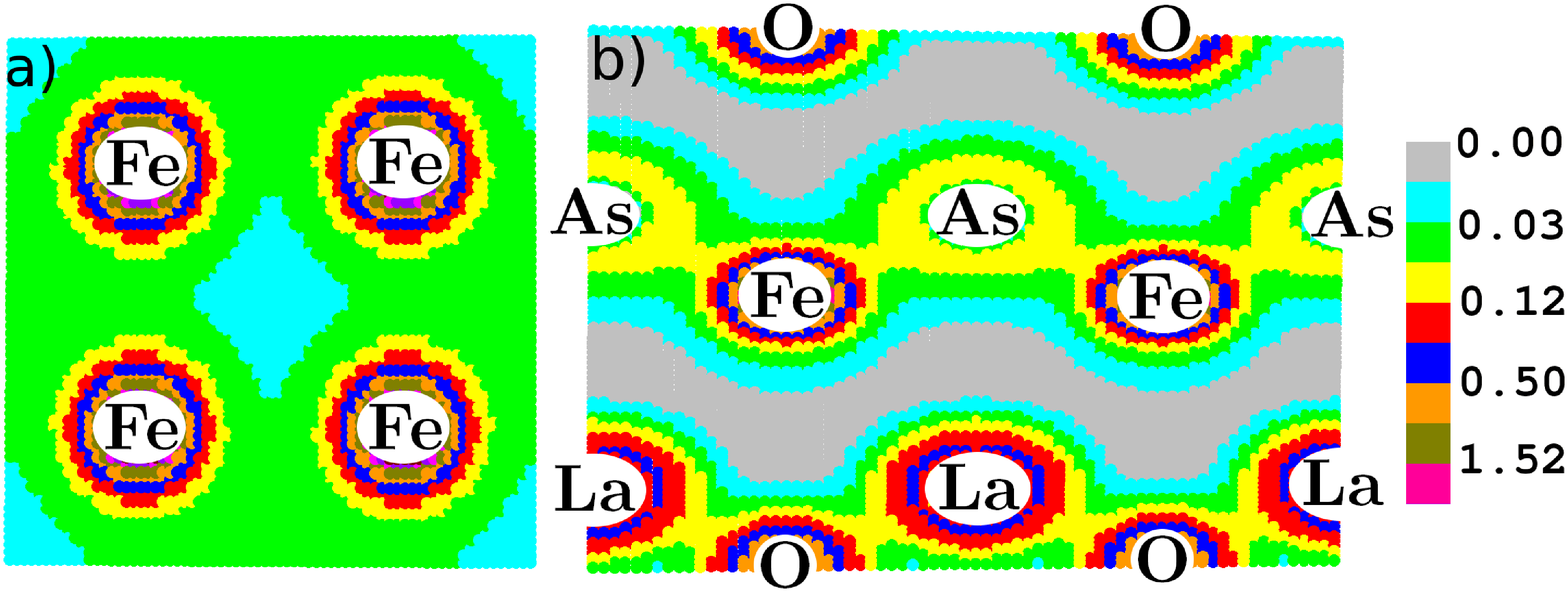}
\caption{(Color online) Charge density distribution of LaOFeAs in
the (001) plane crossing Fe-Fe atoms (a) and in the (110) plane
crossing Fe-As-Fe atoms (b). } \label{fig:fig2}
\end{figure}

The above result indicates that there are competing AFM interactions
between Fe spins. In particular, the AFM coupling between two
next-nearest neighboring Fe spins is very strong. This will
frustrate the spin N\'{e}el structure and give rise to a collinear
ordered AFM ground state.

To explore the origin of these AFM interactions, we have calculated
the charge distribution around Fe and As ions. The result (see Fig.
\ref{fig:fig2}a and \ref{fig:fig2}b) shows that there is almost no
charge distribution between two diagonal Fe atoms, but there is a
strong bonding between Fe and As ions. This indicates that the AFM
coupling $J_2$ is induced by the superexchange bridged by As ions.
This superexchange is AFM because the intermediated state associated
with the virtual hopping bridged by As ions is a spin singlet.

The charge distribution between two nearest Fe ions is finite. Thus
there is a direct exchange interaction between two neighboring Fe
spins. Since there is a strong Hund's coupling between the spins of
3d electrons within each Fe ion, the direct exchange interaction is
found to be ferromagnetic when the distance of two Fe atoms is
between 2.4\AA and 2.85\AA. However, the overall magnetic coupling
$J_1$ between the two nearest Fe spins in LaOFeAs is
antiferromagnetic. Thus $J_1$ is also dominated by the superexchange
interaction bridged by As 4p orbitals.

\begin{figure}
\includegraphics[width=8cm]{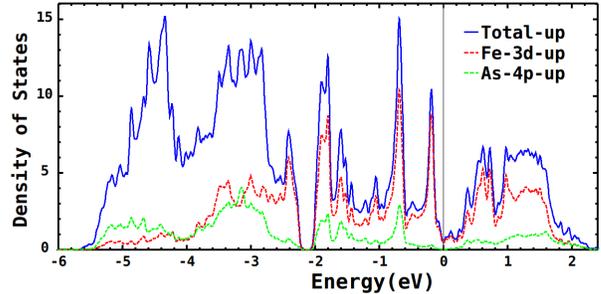}
\caption{(Color online) Total and orbital-resolved partial density
of states (spin-up part) of LaOFeAs in the stripe-ordered
antiferromagnetic state. The Fermi energy is set to zero.}
\label{fig:3}
\end{figure}

In the collinear AFM phase, we find that a small structural
relaxation, for which the lattice is slightly expanded along the
antiferromagnic ordering direction (y-axis in Fig. 1) and shrunk
along the ferromagnetic ordering direction (x-axis in Fig. 1), can
further lower the ground state energy. This changes the angle
between two principal axes in ab-plane, $\gamma$, from $90^\circ$ to
$90.47^{\circ}$. The corresponding energy gain is 7meV. However, we
find that this small lattice distortion affects weakly the band
structure and the Fe moments.

Fig. \ref{fig:3} shows the total and projected density of states of
LaOFeAs at the collinear AFM phase. In comparison with the
nonmagnetic phase, we find that most of the states around the Fermi
level are gapped by the collinear AFM order. This suppresses the
total carrier density by more than two orders of magnitude. The
strong suppression is consistent with the Hall coefficient
measurement which shows that the absolute value of the Hall
coefficient is enhanced by more than 150 times in the AFM phase at
4K versus the nonmagnetic phase above 150K. Furthermore, we find
that the density of states at the Fermi level is also suppressed
compared with the nonmagnetic state\cite{ma}. However, it is not
suppressed as strongly as for the total carrier density. The
corresponding electronic specific heats are evaluated as
0.65mJ/(K$^2\ast$mol) (stripe-ordered AFM) and 4.28mJ/(K$^2\ast$mol)
(nonmagnetic), respectively. This is also consistent with the
specific heat measurement.

We have also calculated the band structure of LaOFeAs with 5\% F- or
5\% Sr-doping by taking the virtual crystal approximation in the
collinear AFM phase. We find that the overall band structure  is
hardly changed by 5\% electron or hole doping. The Fe moment is also
unchanged. Only the Fermi energy moves up or down with electron or
hole doping. However, as shown in Fig. \ref{fig:5} the Fermi surface
changes dramatically.

\begin{figure}
\includegraphics[width=8.5cm]{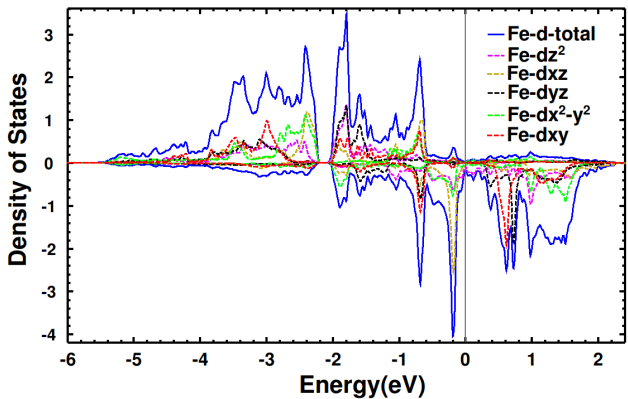}
\caption{(Color online) Total and projected density of states at the
five Fe-$3d$ orbitals around one Fe atom. The Fermi energy is set to
zero.}\label{fig:4}
\end{figure}

By projecting the density of states onto the five 3d orbitals of Fe
(Fig. \ref{fig:4}), we find that the five up-spin orbitals are
almost completely filled and the five down-spin orbitals are only
partially filled. However, the down-spin electrons are nearly
uniformly distributed in these five 3d orbitals. This indicates that
the crystal field splitting imposed by As atoms is relatively small
and the Fe 3d-orbitals hybridize strongly with each other. As the
Hund rule coupling is strong, this would lead to a large magnetic
moment formed around each Fe atom, as found in our calculations. The
frustration between the $J_1$ and $J_2$ terms will suppress strongly
the AFM ordering at the two Fe-sublattices, each connected only by
the $J_2$ terms. This, together with the quantum fluctuation, will
reduce strongly the average magnetic moment around each Fe measured
by experiments.

\begin{figure}
\includegraphics[width=8.5cm]{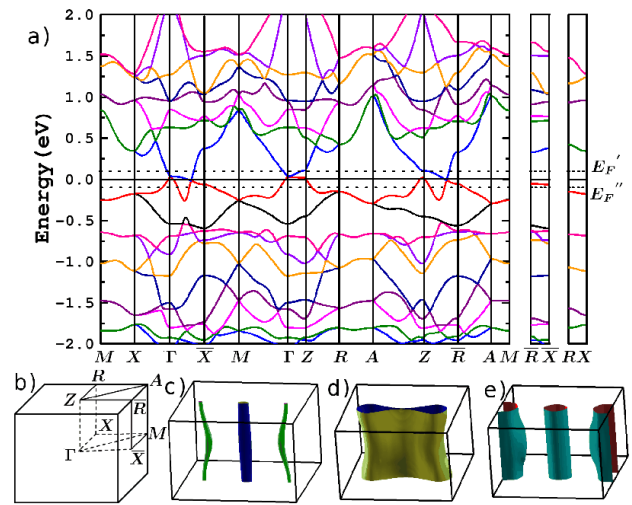}
\caption{(Color online) (a) The electronic band structure of LaOFeAs
in the collinear-ordered antiferromagnetic state with the in-plane
angle $\gamma=90.47^{\circ}$ (see Fig. 1 for the definition of
$\gamma$). The Fermi energy is set to zero. $E_{F}^{\prime}$ and
$E_{F}^{\prime\prime}$ correspond to 5\% F- and 5\% Sr-doping cases,
respectively. (b) The Brillouin zone. (c) The Fermi surface of the
undoped compound: a hole-type cylinder along $\Gamma$Z and two
electron-type pockets between $\Gamma$ and $\bar{X}$. $\Gamma X$
corresponds to the parallel-aligned moment line and $\Gamma \bar{X}$
corresponds to the antiparallel-aligned moment line. (d) The Fermi
surface of the 5\% F-doped compound. (e) The Fermi surface of the
5\% Sr-doped compound.} \label{fig:5}
\end{figure}

Fig. \ref{fig:5} shows the band structure and the Fermi surfaces of
electrons in the collinear AFM state. Unlike in the nonmagnetic
state, there are now only three Fermi surface sheets in undoped
case, one small hole cylinder along $\Gamma$Z and two small electron
pockets formed between $\Gamma$ and $\bar{X}$, crossing the Fermi
level. From the volumes enclosed by these Fermi surfaces, we find
that the hole carrier density is about $1.64\times10^{19}/cm^3$ and
the electron carrier density is about $0.94\times10^{19}/cm^3$. Both
decrease by more than two orders of magnitude in comparison with the
nonmagnetic or square-AFM states\cite{ma}. Upon F(Sr)-doping, the
electron (hole) Fermi surface sheets expand while the hole(electron)
Fermi surface sheets shrink. With 5\% F(Sr)-doping, the whole Fermi
surface becomes electron(hole)-typed and the corresponding
electron(hole) carrier density is $6.31\times10^{20}/cm^3$
($7.01\times10^{20}/cm^3$), increasing by about 25 times compared
with total carrier density in the undoped case.

The above discussion shows that there are strong nearest and next
nearest neighbor superexchange interactions in LaOFeAs. The
interplay between these AFM interactions can affect strongly the
magnetic structure of the ground state. Upon doping, the AFM
ordering will be suppressed. However, we believe that the remanent
AFM fluctuation will survive, similar as in cuprate superconductors.
Thus the effective low energy model for describing these Fe-based
superconductors, no matter whether it is a single- or multi-band
Hamiltonian, should include the frustrated Heisenberg terms defined
by Eq. (\ref{eq:Heisenberg}).

In the Fe-based superconductors, the magnetic fluctuation, induced
by either the AFM superexchange interactions or the on-site Hund's
rule coupling, can be responsible for the superconducting pairing.
The former interaction favors a spin singlet pairing, while the
latter favors a spin triplet pairing. However, the superconductivity
induced by the Hund's rule coupling would generally involve the
interband pairing, which is limited by the available phase space if
the total momentum of Cooper pair is zero. This would suggest that
the spin triplet pairing is not energetically favorable in a system
with strong AFM fluctuations. Moreover, from the study of high-T$_c$
cuprate superconductivity, we know that the strong next-nearest
neighbor AFM interaction favors an extended $s$ or $d_{xy}$-wave
pairing. Therefore, we believe that the leading pairing instability
will be in spin singlet channel, if the superconductivity is driven
by the AFM fluctuation. However, the competing nearest neighbor AFM
interaction may introduce a small new component with different
symmetry, for example a $d_{x^2-y^2}$-wave gap, to the pairing
function. Thus the resulting gap parameter will generally be a
superposition of two components with different symmetries, for
example an extended s plus $d_{x^2-y^2}$ pairing state.

In conclusion, we have presented  first-principles calculations of
the electronic structure of LaOFeAs. We find that there are strong
antiferromagnetic nearest and next-nearest neighbor superexchange
interactions, bridged by As 4p orbitals. The next nearest neighbor
antiferromagnetic coupling is comparable to the nearest neighbor
one. This gives rise to the collinear AFM ordering of Fe spins in
the ground state as observed by neutron scattering. The existence of
strong antiferromagnetic fluctuations in the Fe-based
superconductors bears a strong analogy to the high-T$_c$ cuprates.
This suggests that the superconductivity in these two different
kinds of high-T$_c$ materials may have a common origin.


We wish to thank N.L. Wang, J.L. Luo, P. Dai, Y.H. Su and H.G. Luo
for fruitful discussions. This work is partially supported by
National Natural Science Foundation of China and by National Program
for Basic Research of MOST, China. While this paper was being
finalized, we learnt of the work of S. Ishibashi and coworkers
(arXiv:0804.2963). They obtained a similar band structure.

Note added in revision: In a $2\times 2$ Fe-Fe lattice of the
$J_1$-$J_2$ model with periodic boundary conditions, the $J_1$ and
$J_2$ terms are overcounted by a factor of 2 and 4, respectively.
This overcounting has not been considered in our previous estimation
of $J_1$ and $J_2$, which leads to an overestimation of the values
of these parameters as well as the ratio $J_2/J_1$. This error has
been corrected in this revised version.

\section{Appendix: Determination of $J_1$ and $J_2$}

To determine the value of $J_1$, one need to first evaluate the
energy of a pair of nearest Fe-Fe moments in parallel ($E_{F,1}$)
and anti-parallel ($E_{A,1}$) alignments with respect to a
non-magnetic reference state, respectively. Then from their
difference, one can determine the value of $J_1$ by the following
formula
\begin{equation}
J_1 = (E_{F,1} - E_{A,1})/(2S^2),
\end{equation}
where $S$ is the spin of each Fe ion. It should be emphasized that
$E_{F,1}$ is not necessary to be equal to $-E_{A,1}$ since the
energy of the reference state may not be located exactly at the
middle of the energy between the ferromagnetic and antiferromagnetic
states. This energy lineup between the nonmagnetic and any other
magnetic state needs self-consistent total energy calculations to
determine, that is what we have done. Thus $E_{F,1}$ and $E_{A,1}$
should be determined independent. Similarly, $J_2$ can be determined
from the difference between the energy of a pair of next-nearest
Fe-Fe moments in the parallel ($E_{F,2}$) and anti-parallel
($E_{A,2}$) alignments:
\begin{equation}
J_2 = (E_{F,2} - E_{A,2})/(2S^2).
\end{equation}

\begin{figure}
\includegraphics[width=6cm,height=5cm]{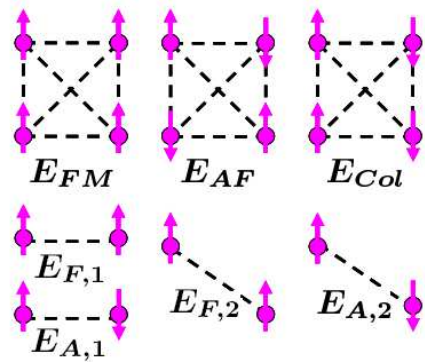}
\caption{The three magnetic configurations and the magnetic bond
energies defined in the text. }\label{Config}
\end{figure}

To determine the values of $E_{F,1} - E_{A,1}$, $E_{F,2}-E_{A,2}$,
we have calculated the total energies of the ferromagnetic
($E_{FM}$), square antiferromagnetic ($E_{AF}$), and collinear
antiferromagnetic ($E_{Col}$) states, respectively. The spin
configurations of these three states are shown in Fig. \ref{Config}.
The corresponding energy differences with respect to the nonmagnetic
state ($E_{NM}$) are (0.0905, -0.10875, -0.21475) eV/Fe. If these
energy differences result mainly from the exchange interactions
between the nearest or next-nearest Fe moments, then we obtain the
following equations
\begin{eqnarray*}
E_{FM}-E_{NM} &= & 2E_{F,1} + 2E_{F,2} = 0.0905 eV, \\
E_{AF}-E_{NM} & =& 2E_{A,1} +2E_{F,2} = -0.10875 eV, \\
E_{Col}-E_{NM}& =& E_{F,1} + E_{A,1}  +2 E_{A,2} = -0.21475 eV.
\end{eqnarray*}
From them, we further find that
\begin{eqnarray*}
E_{F,1} - E_{A,1} & = & 0.0996 eV, \\
E_{F,2}-E_{A,2} & = & 0.1028 eV.
\end{eqnarray*}
Thus the values of $J_1$ and $J_2$ are
\begin{eqnarray}
J_1 & = & 0.0498 eV /S^{2} , \\
J_2 & = & 0.0501 eV/ S^{2} .
\end{eqnarray}

\begin{figure}
\includegraphics[width=8cm,height=6cm]{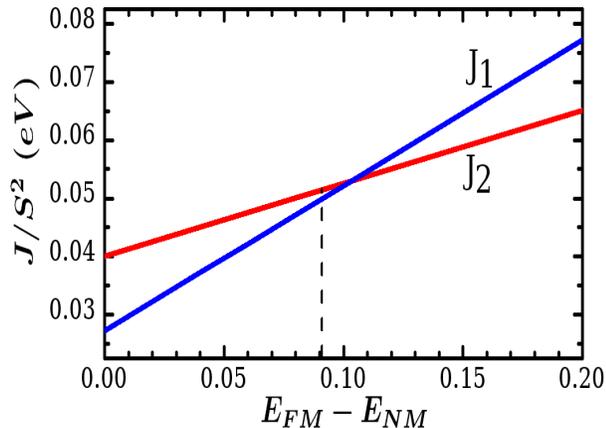}
\caption{Variations of $J_1$ and $J_2$ with $E_{FM}-E_{NM}$. The
dashed line denotes our calculated value.}\label{J-FM}
\end{figure}

The energy of the current ferromagnetic state is less accurately
determined since this state is not a stable state. The error in
$E_{FM}-E_{NM}$ will give rise to the error in $J_1$ and $J_2$. Fig.
\ref{J-FM} shows how $J_1$ and $J_2$ change with $E_{FM}-E_{NM}$. As
we see, we find that $J_2> J_1/2$ even we assume the deviation of
$E_{FM}-E_{NM}$ from our calculated value is as big as 0.1 eV. Thus
we believe that the collinear antiferromagnetic order observed in
LaOFeAs is indeed due to the competition of superexchange
interactions.

The above estimation indicates that $J_1 \sim J_2$ within the error
of calculation for LaOFeAs. In this parameter range, as shown by Yao
and Carlson\cite{yao} with the spin wave approximation, the $J_1$
superexchange interaction term will compete strongly with the
$J_2$-term. This results in a strong reduction of the magnetic
moment of Fe, which would naturally explain why the observed
magnetic moment ($\sim 0.36 \mu_B$) is significantly smaller than
that obtained from the density functional calculations for LaOFeAs.

\end{document}